\documentclass{optica-article}

\journal{opticajournal} 

\articletype{Research Article}

\usepackage{lineno}
\usepackage{xcolor}
\usepackage{multirow}
\usepackage{graphicx}

\begin{document}

\title{Magnetic trapping of an ultracold $^{39}$K-$^{40}$K mixture with a versatile potassium laser system}

\author{Mateusz Bocheński, Jakub Dobosz, Mariusz Semczuk$^*$}
\address{Institute of Experimental Physics, University of Warsaw, Pasteura 5, 02-093 Warsaw, Poland}

\email{\authormark{*}msemczuk@fuw.edu.pl}

\begin{abstract*} 
We present a dual isotope magneto-optical trap (MOT), simultaneous sub-Doppler laser cooling, and magnetic trapping of a spin-polarized $^{39}$K-$^{40}$K Bose-Fermi mixture realized in a single-chamber setup with an unenriched potassium dispenser as the source of atoms. We are able to magnetically confine more than $2.2\times10^5$ fermions ($F=9/2\,m_F=9/2$) and $1.4\times10^7$ bosons ($F=2\,m_F=2$) with a lifetime exceeding 1.2~s. For this work, we have developed a versatile laser tailored for sub-Doppler cooling of all naturally occurring potassium isotopes and their mixtures. This laser system incorporates innovative features, such as the capability to select an isotope by activating or deactivating specific acousto-optic modulators that control the light seeding tapered amplifiers. Switching between isotopes takes $\sim$1\textmu s without any mechanical adjustment of the components. As a final step in characterizing the laser system, we demonstrate sub-Doppler cooling of $^{41}$K.

\end{abstract*}

\section{Introduction}

The research of ultracold isotopic mixtures is of significant importance for diverse scientific investigations. Although some rationales for delving into this domain are somewhat technical in nature, such as refining the production techniques for fermionic $^3$He$^*$~\cite{PhysRevLett.97.080404}, $^6$Li~\cite{doi:10.1126/science.1059318}, ${}^{40}$K~\cite{wu2011strongly}, and bosonic ${}^{39}$K~\cite{cabrera2018quantumPhD} and ${}^{85}$Rb~\cite{PhysRevLett.101.040402} quantum gases through sympathetic cooling, the implications extend far beyond mere technical advancements. Isotopic mixtures are advantageous from an experimental standpoint for at least three reasons: the almost identical mass of constituents limits their gravity-induced separations in shallow traps used in the quantum degenerate regime, they experience nearly identical confining optical potential, and the laser systems can operate at nearly the same wavelength avoiding the use of achromatic or broadband-coated optics. Ultracold mass-imbalanced mixtures would separate due to gravity, which needs to be compensated by magnetic levitation~\cite{Lercher2011} or by adding a confining optical potential at a suitable wavelength~\cite{PhysRevLett.119.233401}. These are largely the reasons why isotopic mixtures were used in the first successful demonstrations of the simultaneous superfluidity of fermionic and bosonic systems~\cite{doi:10.1126/science.1255380} and the miscibility of quantum liquids \cite{PhysRevLett.101.040402}. Photoassociation studies of homo and heteronuclear mixtures of nearly identical mass~\cite{PhysRevA.66.061403} can also provide valuable information about the mass scaling of molecular potential models or the Born-Oppenheimer approximation. 

The availability of three stable isotopes, bosonic ${}^{39}$K and ${}^{41}$K, and fermionic ${}^{40}$K, makes potassium a versatile tool in modern ultracold experiments~\cite{PhysRevLett.123.233402,Pauli_blocking_40K}. All isotopes have already been brought to the quantum degenerate regime both by direct evaporation~\cite{PhysRevA.86.033421,DeMarco_DFG,PhysRevA.79.031602} and by sympathetic cooling by other atomic species~\cite{PhysRevLett.99.010403,aubin2006rapid,modugno_41K_BEC}. With isotopic mixtures, off-resonance excitation of transistions in one isotope while cooling the other one (and vice versa) could limit the efficiency of sub-Doppler cooling. However, with the proper adjustment of the experimental parameters, it was possible to simultaneously cool $^3$He$^*$ and $^4$He$^*$~\cite{PhysRevLett.97.080404}, $^6$Li and $^7$Li as well as ${}^{39}$K with ${}^{41}$K~\cite{cabrera2018quantumPhD}, and ${}^{40}$K with ${}^{41}$K~\cite{wu2011strongly}. In the latter two cases, ${}^{41}$ K was used as a coolant to bring the mixture to quantum degeneracy. 

This work provides the first demonstration of the so far experimentally unexplored ultracold mixture of ${}^{39}$K and ${}^{40}$K which we cool to sub-Doppler temperatures (below 30~\textmu K), spin polarize and magnetically trap. It could be considered the first step towards realizing a proposal of Bazak and Petrov~\cite{bazak2018stable}, where they consider a composite fermionic ${}^{39}$K${}^{40}$K molecules in a quasi-two-dimensional confinement as a promising, and thanks to our work realistically accessible, candidate to obtain a strongly $p$-wave-attractive Fermi gas.

Our experimental setup uses unenriched alkali dispensers located 7~cm away from the trapping region, making it relatively compact. The lifetime of the magnetically trapped mixture is on the order of 1.2~s, the result of heating potassium dispensers to maximize the number of ${}^{40}$K atoms, which also increases the number of ${}^{39}$K atoms in the background gas. With enrichment reaching 3\%, the dispensers could be operated at a much lower temperature, comparable to what we typically use when working only with ${}^{39}$K or ${}^{41}$K, which would extend the background-limited lifetime to nearly 10~s. Even though the lifetimes reported here are short for conventional evaporation, the initial atom number is sufficiently large such that further evaporation could be performed after moving the clouds to a second chamber with much better vacuum. Thus, our approach could be an alternative to using 2D MOTs or Zeeman slowers. It is worth noting that for many experiments performed in optical tweezers or for photoassocation spectroscopy such a short lifetime might be sufficient. 

An important aspect of this work is the demonstration of a novel approach to constructing laser systems for laser cooling of multiple isotopes. Our design is modular, allows for a 1~\textmu s-scale switching between isotopes, has built-in coherence between the cooling and the repumping beam, and, most importantly, can be used for simultaneous sub-Doppler cooling of isotopic mixtures, as shown here. Mechanical adjustments are not necessary, and the extension to other potassium isotopes such as $^{37}$ K and $^{38m}$ K requires the addition of only a pair of acousto-optic modulators per isotope. We provide a thorough characterization of all modules comprising the laser system validating it with sub-Doppler cooling of ${}^{41}$K to 11~\textmu K. In conjunction with our previous works on ${}^{39}$K~\cite{dobosz2021bidirectional} and ${}^{40}$K~\cite{bochenski2024sub} we show that the discussed laser system enables state-of-the-art cooling of all potassium isotopes individually and as isotopic mixtures. 

\section{Laser system design}

Each laser cooling experiment depends on the capability to adjust the laser beam frequency to optimize the mechanisms suitable for a specific cooling phase. Although electro-optic modulators are sometimes used for this purpose, the most common method is the utilization of acousto-optic modulators (AOMs) in a double-pass setup~\cite{10.1063/1.1930095}. The latter method avoids introducing undesired frequency components into the experiment and can also serve as a fast shutter, crucial for promptly turning off the laser light. Nevertheless, there is a fundamental limitation to this approach: achieving close to 90\% diffraction into the desired order is rare, resulting in a more than 20\% loss of laser power in a double-pass configuration, with actual experimental losses often closer to 40\%. Beam-shape distortions are frequently observed, which may lead to problems with coupling efficiency when transmitting light through fibers. Consequently, the design of a laser system must satisfy numerous, sometimes conflicting requirements. Increasing the laser power to compensate for losses is not always feasible, as typical high-power sources utilize tapered amplifiers that exhibit output powers closely tied to the wavelengths employed. While cooling cesium or potassium does not present significant power-related challenges if the cost of the laser system is a secondary issue, experiments with lithium or sodium face more noticeable technical limitations in terms of the laser power that can be utilized.

In tapered amplifiers, several tens of milliwatts of seed power can be amplified up to several watts within a broad wavelength range, often spanning 10-20~nm around the specified central wavelength. As a result, multiple frequency components, even separated by several THz, can be simultaneously amplified~\cite{valenzuela2013multiple}. Unfortunately, not all that power is useful, because the spatial quality of the output beam is usually low, with M$^2\sim1.5-2$. This results in a low efficiency of coupling this light into a single-mode fiber for distribution within the experimental setup. 

We have developed a modular design for the laser system, where modules are interconnected through single-mode, polarization-maintaining fibers. This approach facilitates the optimization of each module's performance independently without causing misalignment in the overall system. The laser system comprises three modules, which are detailed in the subsequent sections: the stabilization module (SM) which generates frequency-stabilized light on the $D_1$ and $D_2$ lines, the frequency tuning module (FM) where fine-tuning of frequency and selection of the cooled isotope is carried out using acousto-optic modulators, and the amplification module where the final beam preparations take place, power is amplified using tapered amplifiers, and the output is directed to the experimental setup through fibers.

The structure of the energy levels in potassium exhibits certain peculiarities, such as a small splitting of the excited $P_{3/2}$ state and the inverted hyperfine structure of the ground state of $^{40}$K. These features make the convention of naming transitions as 'cooling' or'repumping' somewhat superficial. In order to maintain consistency with the existing literature, we have opted to adhere to the conventional terminology in the article. For bosonic and fermionic isotopes, the transitions on the $D_2$ ($D_1$) line identified as 'cooling transitions' are $^2S_{1/2}\,F=2 \to {}^2P_{3/2},F'=3$ ($^2S_{1/2}\,F=2 \to {}^2P_{1/2},F'=2$) and $^2S_{1/2}\,F=9/2 \to {}^2P_{3/2}\,F'=11/2$ ($^2S_{1/2}\,F=9/2 \to {}^2P_{1/2}\,F'=9/2$), while the term 'repumping transitions' refers to $^2S_{1/2}\,F=1 \to {}^2P_{3/2}\,F'=2$ ($^2S_{1/2}\,F=1 \to {}^2P_{1/2}\,F'=1$) and $^2S_{1/2}\,F=7/2 \to {}^2P_{3/2}\,F'=9/2$ ($^2S_{1/2}\,F=7/2 \to {}^2P_{1/2}\,F'=7/2$). In this context, transitions with fractional total angular momentum $F$ correspond to $^{40}$K.

\subsection{Stabilization module (SM)}

\begin{figure}[htbp]
\centering\includegraphics[width=.9\textwidth]{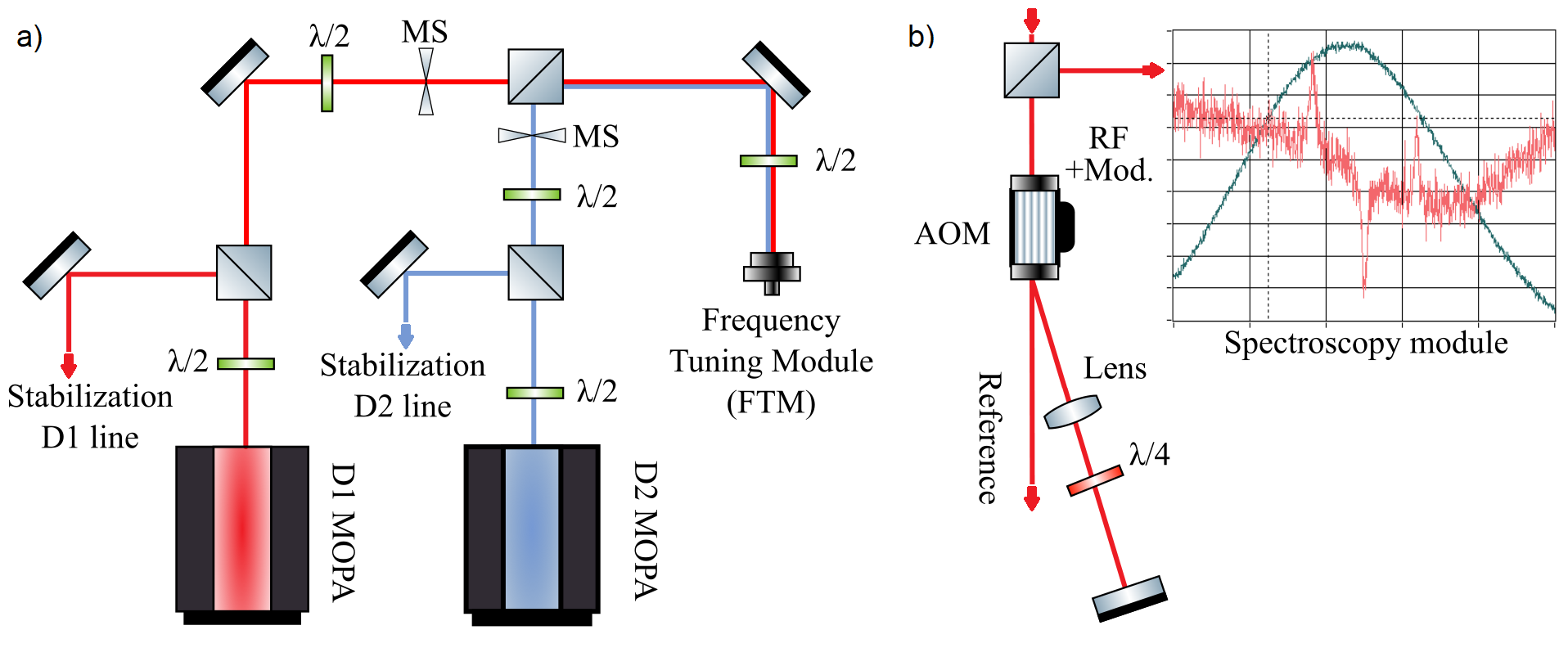}
\caption{a) Layout of the stabilization module. Beams from two MOPAs (Master Oscillator Power Amplifier)) stabilized on the $D_2$ or $D_1$ transitions in $^{39}$K are overlapped on a polarizing beam splitting cube (PBS) and sent to the frequency tuning module. The $D$ line that is used at any given cooling stage is selected by opening or closing an appropriate mechanical shutter (MS); b) the section responsible for frequency stabilization. The layouts for both $D$ lines are nearly identical. The light from the tapered amplifier passes through the acousto-optic modulator (AOM) in a double-pass configuration and is sent to the potassium vapor cell in the spectroscopy module. The central frequency of the AOM is modulated to derive an error signal from Doppler-free saturated absorption spectroscopy (shown for the $D_2$ line in the inset). The 0th-order beam can be used as a frequency reference or for diagnostics.}
\label{SM}
\end{figure}

The main function of this module (see Figure~\ref{SM}) is to provide a sufficiently high power of frequency-stabilized light to seed tapered amplifiers in the amplification module after frequency shifting in the frequency tuning module. In our implementation of the laser system, the module generates light at frequencies $f_\mathrm{D1} = 389.286144$ THz for the $D_1$ line cooling and $f_\mathrm{D2} = 391.016170$ for the $D_2$ line cooling. The setup comprises two MOPA systems that contain an ECDL and a tapered amplifier (Toptica Photonics TA Pro). One of the MOPAs is optimized by the manufacturer for potassium cooling at 767~nm, whereas the other was originally purchased for rubidium (optimized for 780~nm) but is currently being used for the potassium $D_1$ line at 770~nm, although with a minor decrease in output power. Doppler-free saturated absorption spectroscopy of $^{39}$K performed in a commercial module CoSy (TEM Messtechnik GmbH) is used to generate the error signal for frequency stabilization. Initially, this signal was obtained with lock-in detection and current modulation of ECDLs. This, in turn, caused the modulation of the frequency of the cooling light that was a source of instabilities in the number of atoms and limited the performance of $D_1$ gray molasses. We therefore implemented an alternative approach where the light sent to the CoSy module is first frequency shifted by an acousto-optic modulator in a double-pass configuration. We use $^2 S_{1 / 2}, F= 2 \to ^2P_{1 / 2}, F'=2$ and $^2 S_{1 / 2}, F= 2 \to ^2P_{3 / 2}, F'$ transitions for stabilizing the lasers operating on the $D_1$ and the $D_2$ line, respectively. The hyperfine structure of $^2P_{3 / 2}$ is not resolved; therefore, we assign the excited state a generalized total angular momentum $F'$ \cite{PhysRevA.70.052505}. With an external VCO we could drive the AOM at 125~MHz (101~MHz) and modulate it by $\pm$3 MHz at 7~kHz (15~kHz) for the $D_2$($D_1$)-line stabilization using a home-built bias tee. This way the lock-in detection could still be implemented with the laser controller while removing the fluctuation of the frequency of the cooling light. This simple modification increased the number of $^{39}$K atoms trapped in a magneto-optical trap by a factor of two and turned out to be essential to achieve state-of-the-art sub-Doppler cooling in the setup \cite{bochenski2024sub, dobosz2021bidirectional}. Our experience, including generating the error signal with a modulation-free spectroscopy setup, makes us conclude that the exact method of stabilizing the frequency does not seem to matter as long as it does not lead to the frequency modulation of the cooling light.

The MOPA systems generate $\sim$1~W of light each. The beams are then overlapped on a polarizing beam splitting cube and coupled into a single, single mode polarization-maintaining fiber with efficiency of $\sim$60\%. Care has been taken to ensure that the beams are coupled into perpendicular axes of the fiber to minimize polarization drifts. Each beam path can be independently blocked with a mechanical shutter (Uniblitz LS6Z2) which allows us to choose whether the $D_1$-line light or the $D_2$-line light is sent to the next module (see  Fig.~\ref{SM}). Switching between $D$ lines takes about 1~ms, with a 1.6~ms delay after applying the trigger signal. With the current shutter driver it takes $\sim$12~ms before a shutter that has been closed could be opened again. This imposes some limitations if one would like to turn on the $D_2$-line light immediately after $D_1$ line cooling, but in most applications we do not find it to be a serious constraint. Moreover, it can be easily overcome with a faster shutter~\cite{bowden_shutter}.

\subsection{Frequency tuning module (FTM)}

The entire module is built on an aluminum plate ($90\times 60\times 2$~cm). Most aluminum pedestals that we use to mount optical components are attached to the plate using epoxy. Standard laboratory clamps are utilized to secure supports for lenses, acousto-optic modulators, and irises, with M6 holes being drilled and threaded at designated positions. The light is transmitted from the SM using a single-mode, polarization-maintaining fiber, where the $D_1$ and $D_2$ light propagate along two primary orthogonal axes of the fiber. By employing a half-wave plate and a polarizing beam splitting cube we ensure that the power from each axis of the fiber is equally distributed into two distinct paths.

\begin{figure}[htbp]
\centering\includegraphics[width=\textwidth]{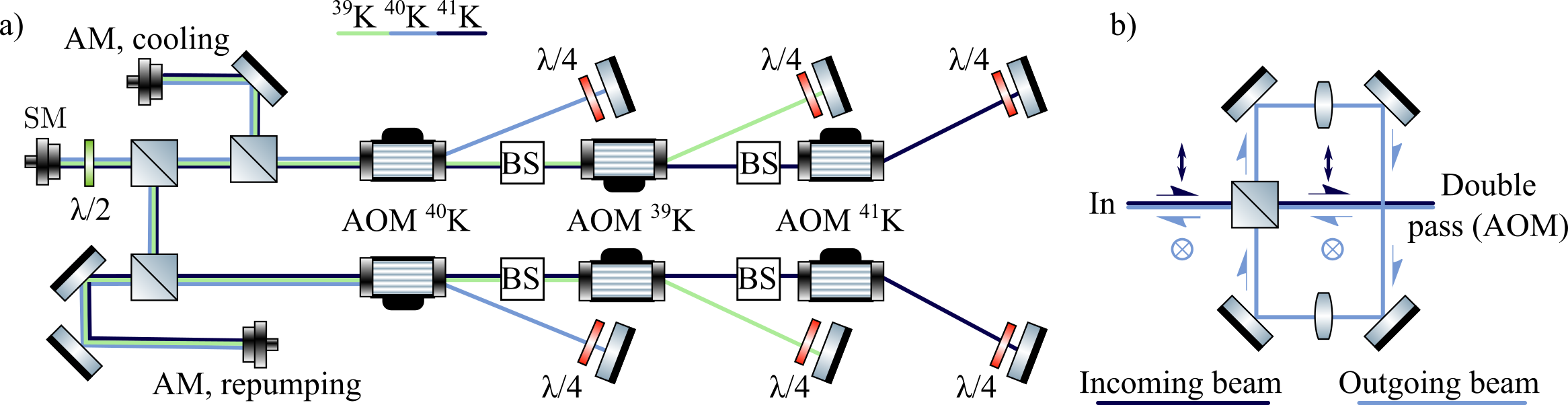}
    \caption{a) Layout of the frequency tuning module. Light originating from the stabilization module (SM) is split into two separate paths to address the cooling and repumping transitions. Three modulators in a double-pass configuration are utilized in each path to produce the necessary frequencies for the three stable potassium isotopes. Subsequently, the light from each path is guided into a single-mode, polarization-maintaining fiber and then sent to the amplification module, identified as 'AM, cooling' and 'AM, repumping'. The abbreviation 'BS' refers to the beam shaping section illustrated in b), where the incoming beam could be adjusted using lenses to improve the diffraction efficiency of the acousto-optic modulator (AOM); we have not taken advantage of this option. After the beam undergoes a double-pass and polarization rotation, it is reflected by the polarizing beam splitting cube (PBS). Subsequently, its shape is fine-tuned using lenses to improve coupling into the optical fiber before being redirected to the opposite side of the PBS. Mirrors are used to fine-tune fiber coupling. This method enables the individual beam shaping of light for each isotope. To improve clarity, the incoming and outgoing beams are maintained apart.}
    \label{FTM}
    \end{figure}

In each path, there are three acousto-optic modulators set up in a double-pass configuration, positioned consecutively so that the 0th order beam of one AOM serves as the input beam for the next (refer to Figure~\ref{FTM}a). The diffracted beams (either 1st or -1st order, depending on the AOM) undergo two passes through the AOMs and are coupled into a shared single-mode, polarization-maintaining fiber that transmits light to the amplification module. We have chosen to use AOMs that operate at central frequencies of 80~MHz, 110~MHz, and 350~MHz, as these are readily available from various manufacturers and can achieve a diffraction efficiency of more than 85\% in a single pass. The selected operating frequencies of the modulators consider the inclusion of single-pass 80~MHz AOMs at the output of the amplification module (see \ref{lab:AM} for a discussion), which are used to control the intensity of the cooling beams and serve as fast shutters. Consequently, the frequency difference between the light produced by the cooling and repumping paths does not correspond to the hyperfine splittings of potassium isotopes.

After the frequency shift with the AOMs is performed, the light from each path is coupled to a dedicated fiber and is sent to the amplification module. Under normal circumstances, it is impossible to ensure optimal diffraction efficiency of each AOM and efficient coupling into a fiber because the laser beam travels a distance of up to 2~m (for the last AOM in the chain) and diffraction becomes non-negligible, making it nearly impossible to assure satisfactory performance for all AOMs in the chain but the first one. We have developed a systematic approach that mitigates the issue of diffraction and technical requirements imposed by modulators (like the optimal beam size). We optimize the diffraction efficiency of the first modulator configured in the double-pass configuration, then we optimize the coupling efficiency into the fiber by appropriately shaping the beam with lenses. We add another modulator and shape the 0th order beam from the previous AOM with a set of lenses to optimize its diffraction efficiency. It results in a very poor fiber-coupling efficiency on its own. To improve that, we exploit a key feature of modulators which are set up in a double-pass configuration: the frequency-shifted beam has a perpendicular polarization to the incoming beam due to the use of a quarter-wave plate. In our approach (see Figure~\ref{FTM}b) the beam enters an AOM after passing through a polarizing beam-splitting cube, then passes the AOM twice, and on its way back it reflects off the PBS because its polarization is now perpendicular to the polarization of the incoming beam. It is then rerouted to the opposite wall of the PBS with the help of four mirrors. The space between the mirrors is used to shape the beam, and each subsequent beam-shaping stage must consider the presence of the beam-shaping optics in all previous stages. This is somewhat tedious to set up for the first time, but allows us to obtain >50\% fiber coupling efficiency for the last modulator in the chain (here, for $^{41}$K), while for the first one (for $^{40}$K) it approaches 75\%. The difference is primarily the result of the beam distortion caused by the crystals in the AOMs.

The effective frequency tuning range of the entire laser system is related to the bandwidth of AOMs, but in a somewhat indirect manner: the frequency can be changed as long as the power delivered to the amplification module is sufficiently high for the tapered amplifiers to operate without noticeable amplified spontaneous emission. This minimum power level, 20~mW for the repumping amplifier and 10~mW for the cooling amplifier, has been determined by observing the emission spectra of the amplifiers using an Ocean Optics USB4000 spectrometer.

The modular design of the laser system enables its extension to other potassium isotopes by using the 0th order of the last acousto-optic modulator in each path of the FTM. Based on the hyperfine structure values of radioactive potassium isotopes reported in~\cite{williamson1997magneto}, we have calculated that detunings of -545~MHz and -476~MHz (-756~MHz and 449~MHz) from the cooling and repumping transitions are required, respectively, to address transitions in $\mathrm{^{37}}$K ($\mathrm{^{38}}$K). Such detunings can be easily obtained by commonly available AOMs set up in double-pass configurations, with central frequencies of 250~MHz and 350~MHz.

\subsection{Amplification module (AM)\label{lab:AM}}

\begin{figure}[htbp]
\centering\includegraphics[width=\textwidth]{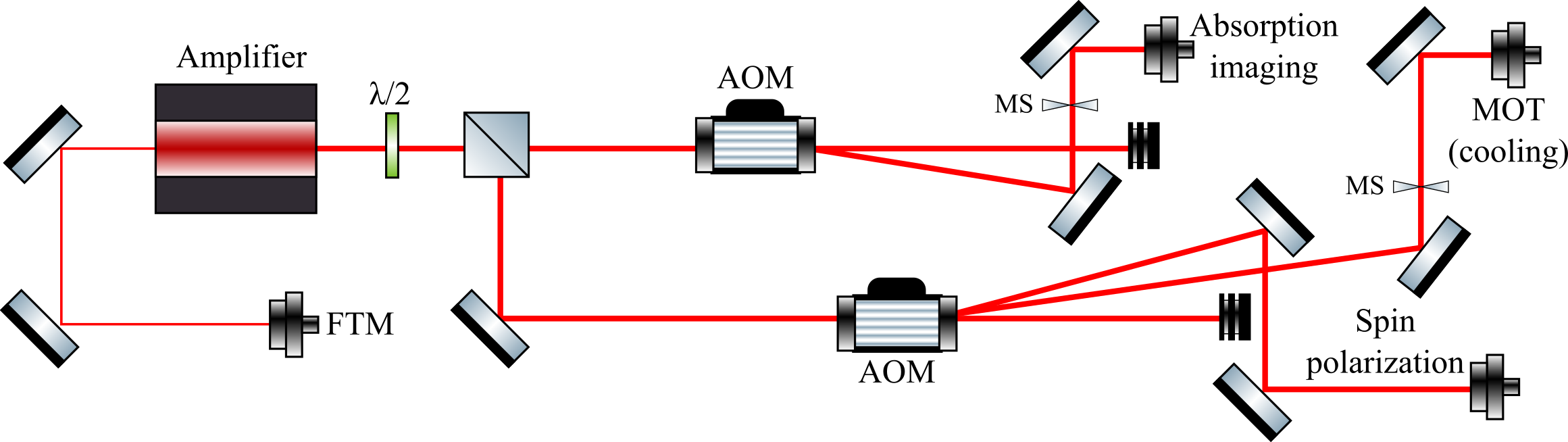}
\caption{Layout of the the cooling section of the amplification module. Light from the frequency tuning module (FTM) seeds the optical amplifier which output is then split into two paths. In the first path, an acousto-optic modulator controls the absorption imaging. The modulator in the second path is used to select between the cooling path for the magneto-optical trap and the spin polarization path.}
\label{AM}
\end{figure}

The amplification module, illustrated in Figure~\ref{AM}, comprises two tapered amplifiers: one generates cooling light (''cooler'') and the other repumping light (''repumper''). Both amplifiers are seeded with light from the frequency tuning module.  The amplification module is divided into distinct sections for repumping and cooling light. The key difference between these segments lies in the absence of an absorption imaging path in the repumping section and the use of AOMs with different diffraction orders (-1st and +1st for the cooling and repumping sections, respectively). The light emitted from the ''cooler'' (''repumper'') is subjected to a frequency change of -80~MHz (+80~MHz) after passing through an Acousto-Optic Modulator (AOM) before being coupled into a dedicated single-mode polarization-maintaining fiber. This fiber then transmits the light to another optical table where the vacuum chamber is located. Acting as fast switches, the modulators also facilitate the selection of the fiber to which the light is directed: by changing the frequency of the rf signal driving the modulator we can select if the light is used for cooling/repumping or optical pumping/spin polarization. This approach works well because the power needed for optical pumping is relatively low and as a result we drive the AOM 25~MHz below the central frequency. Even then more than 50~mW can be deliverd to the vacuum chamber. Driving 25~MHz above the central frequency is used for the idle mode, which ensures that the AOM is kept on also when no cooling or optical pumping beams are in use, thus mitigating thermal effects that occur in the modulator. The frequency shift that has been introduced is offset by the FTM. The absorption imaging beam is controlled with a separate AOM.

The amplification module employs two MOPA systems (Toptica TA PRO) identical to those utilized in the stabilization module. By eliminating the mirrors between the ECDL laser diodes and the amplifiers, we were able to introduce external light from the frequency tuning module to seed the amplifiers.

Finally, mechanical shutters (Uniblitz LS6Z2) are used to prevent light leakage into the system. It has been observed that even when the acousto-optic modulators are switched off, there is a light leakage at the microwatt level that can negatively impact the lifetime of atoms held in either a magnetic or optical dipole trap.

The system is designed to obtain the highest power for cooling and repumping light, limited only by the maximum power of the amplifiers and the efficiency of the AOM in the amplification module. For all isotopes, the powers available close to the vacuum chamber are within the range of 300-500~mW (200-400~mW) for the cooling (repumping) light. The nominal output power of the tapered amplifiers is 1.5~W, but we have observed a deterioration of amplification over time. However, the main sources of apparent loss are the dependence of seeding power on the isotope due to losses in the FTM and the 50\% coupling efficiency into optical fibers connecting the AM and the vacuum chamber. Rather low fiber coupling efficiency is typical for light from tapered amplifiers as they tend to emit light with rather poor spatial quality, with M$^2\sim1.5-2$.

\subsection{Switching between isotopes - timing}

The major advantage of the laser system that distinguishes it from known designs is its ability to remotely and quickly switch between isotopes without the need for any mechanical adjustments. This would be very useful for experiments that require a fast repetition rate~\cite{farkas2010compact}. On a practical note, the ability to switch to $^{39}$ K when working primarily with other potassium isotopes has proven to be extremely useful when debugging or optimizing the performance of the experimental setup.

\begin{figure}[htbp]
\centering\includegraphics[width=\textwidth]{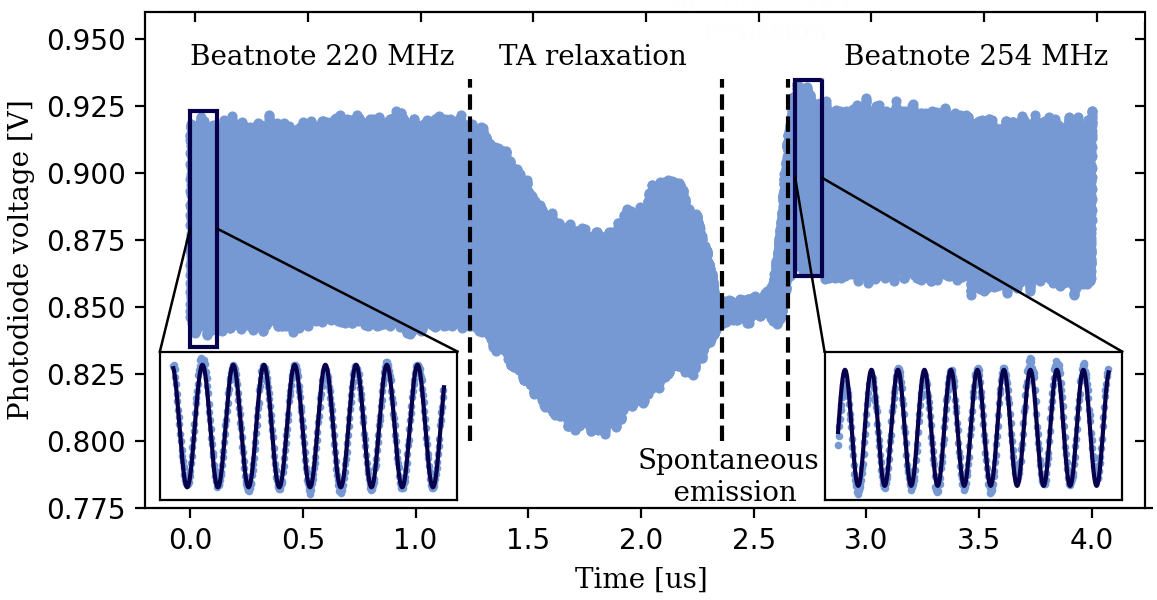}
\caption{The beatnote signal indicating switching time between frequencies for laser cooling different isotopes successively. The figure displays four distinct regions delineated by dashed vertical lines, representing the beatnote between the stabilized laser and the repumping beam for cooling $^{39}$K, TA relaxation, TA spontaneous emission (without a discernible beatnote), and the beatnote following the activation of the repumping beam for the $^{41}$K isotope. In the inset, an amplified beat signal (light blue) is depicted, with the fitting of a sine function (navy blue).}
\label{switching_time}
\end{figure}

To determine the switching time between isotopes, we have used an oscilloscope to record the beatnote between the reference light from the cooling amplifier and the light from the repumping amplifier seeded with the $^{41}$K repumper (\textit{Beatnote 220~MHz} region in Figure~\ref{switching_time} and the associated inset). Then we turn off the $^{41}$K AOM and turn on the AOM responsible for the $^{39}$K repumper (\textit{Beatnote 254~MHz} region in Figure~\ref{switching_time} and the associated inset) using the same experimental sequence that we would use in the actual experiment. Figure~\ref{switching_time} shows four distinct regions that appear in such a measurement. The first and last regions correspond to repumping amplifiers seeded with $^{41}$ K and $^{39}$ K light, respectively. We have confirmed the expected beatnote frequencies by fitting sinusoidal functions to both beatnote regions. After the $^{41}$ K is turned off there is a time (\textit{TA relaxation}) in which the intensity of light decreases, but the repumping amplifier operates at the initial frequency. The next region shows spontaneous emission of the amplifier where no beatnote is observed, and the intensity of the photodiode signal drops. The turn-off time of the AOMs in the frequency tuning module is below 200~ns therefore the relaxation of the amplifier does not seem to be related to the presence of the seeding light. On the basis of multiple measurements, we have estimated the switching time to be on the order of 1.5~\textmu s. This coincides with the timing resolution of our experimental control system and at this point in time we are unable to confirm if it is possible to have a significantly shorter switching time and if the relaxation of the amplifier plays an important role on a shorter time scale.

\section{Sub-Doppler cooling of $^{41}$K}

The discussed laser system has been used to demonstrate state-of-the-art sub-Doppler cooling of $^{39}$K~\cite{dobosz2021bidirectional} and $^{40}$K~\cite{bochenski2024sub}. For completeness, here we present sub-Doppler cooling of the remaining stable isotope, $^{41}$K. We use the same single-chamber apparatus in which cooling of other isotopes has been performed. Three retroreflected trapping beams with a diameter of about 25~mm and intensities of 23$I_\mathrm{s}$ and 17.5$I_\mathrm{s}$ for the cooling and the repumping beam, respectively, create a magneto-optical trap (MOT). With a magnetic field gradient of 7~G/cm, a 6.2~$\mathrm{\Gamma}$ detuning of the cooling and repumping beams from their respective transitions, and a 6~s loading time the number of atoms in the MOT saturates at $N = 5\times 10^7$ with a temperature of $\sim$10~mK. This corresponds to the phase-space density of $\mathrm{\rho \approx 10^{-11}}$. Here,  $\Gamma = 6.035$~MHz is the natural linewidth of the $D_2$ line, and $I_s = 1.75$~mW/cm$^2$ is the saturation intensity~\cite{tiecke2010properties}. The strongest magnetic field gradient is in the horizontal direction due to the arrangement of the MOT coil. However, throughout the article, we quote the magnetic field gradient in the direction of gravity (vertical direction), which is half the horizontal value (14~G/cm during MOT loading). Atoms are released from a dispenser containing a natural isotope mixture (approximately 6.73 \% of $^{41}$K), placed 7~cm from the center of the trap. The cooling performance is characterized when the dispenser current is 4~A to improve the SNR. On a daily basis, the dispenser is turned off or heated resistively with a current of 3.6~A.

Mechanical shutters in the spectroscopy module limit the switching time from the $D_2$ line to the $D_1$ line to approximately 1.8 ms, with a jitter of around 0.8~ms. During this time, the cloud falls freely and expands, decreasing its density. This is a critical issue because, at this stage, the atoms are relatively hot compared to the MOTs of other species. The conventional D1-D2 compressed MOT (D1-D2 CMOT) method~\cite{chen2016production} cannot be applied when compressing the cloud as it requires addressing the repumping transition on the $D_2$ line and the cooling transition on the $D_1$ line which is impossible with the design of our laser system. However, it is possible to perform cloud compression and cooling in a modified D1-D2 CMOT by amplifying the light for both $D$ lines simultaneously in the amplification module.

Compression in the modified D1-D2 CMOT begins immediately after the MOT has reached its steady-state atom number. We raise the magnetic field to 35~G/cm and set the $D_2$ cooling (repumping) light detuning at -3$\Gamma$ (-3.5$\Gamma$) and intensity at 11.5$I\mathrm{_s}$ (9$I\mathrm{_s}$). Simultaneously, we open the shutter located in the spectroscopy module sending the $D_1$ line light to the amplification module. This results in a detuning of +4.5$\Gamma$ (+7.5$\Gamma$) and an intensity of 7.3$I\mathrm{_s}$ (7.3$I\mathrm{_s}$) for the cooling (repumping) $D_1$ light. The temperature decreases to 1~mK and the phase-space density increases to $\mathrm{\rho\approx 10^{-9}}$, without significant loss of atoms. The D1-D2 CMOT is crucial for enhancing the cloud's density and facilitating its cooling. Furthermore, it plays a critical role in allowing the shutters a sufficient time (2.6 ms) to transition between the $D_2$ and $D_1$ lines.

\begin{figure}[htbp]
\centering\includegraphics[width=\textwidth]{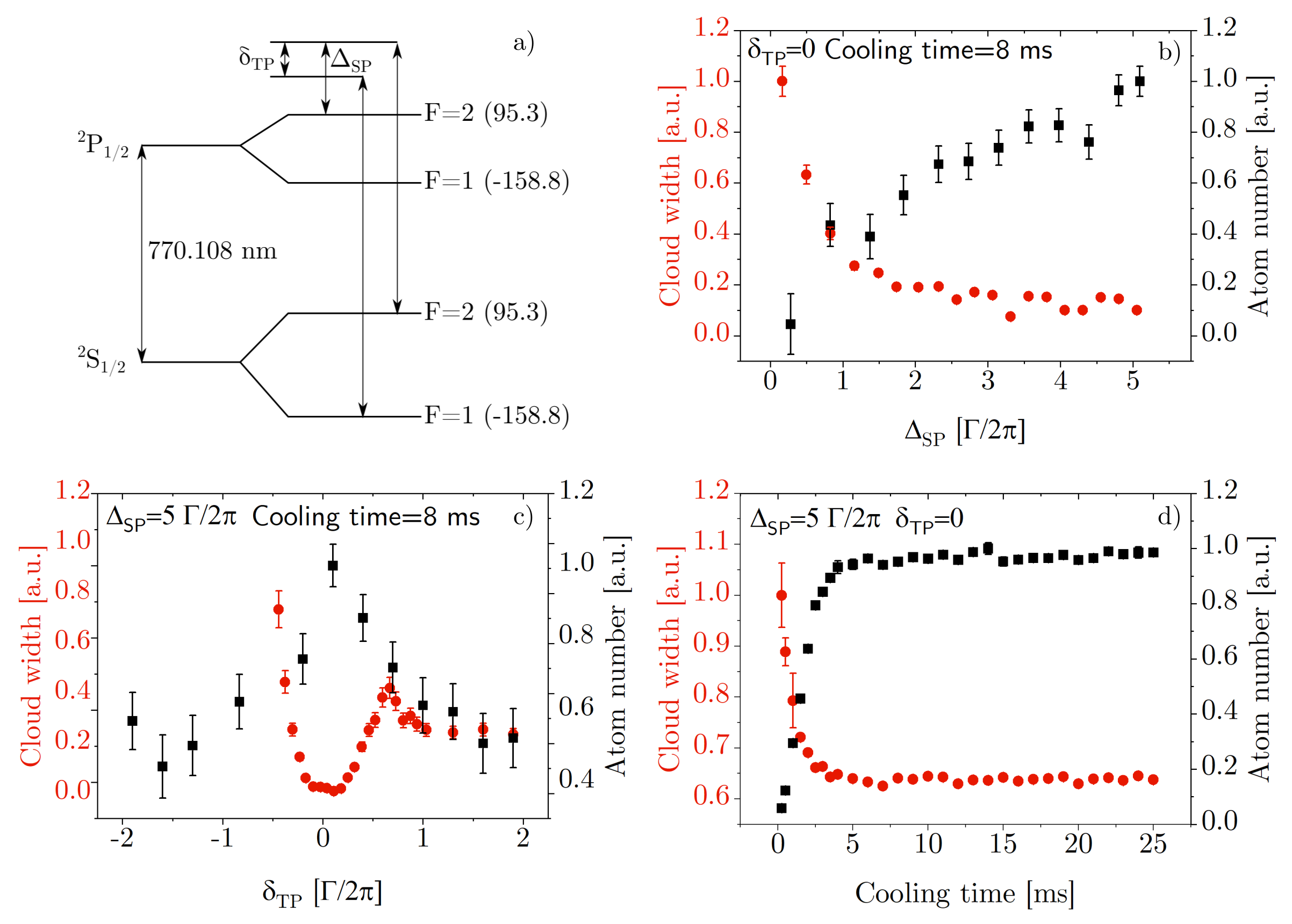}
\caption{a) Hyperfine structure of the $D_1$ line of $^{41}$K with frequencies of the cooling and repuming light used in gray-molasses cooling. Two-photon, and single-photon detunings are labeled as $\delta_{\mathrm{TP}}$ and $\Delta_{\mathrm{SP}}$, respectively. The hyperfine shifts are in units of MHz. b)-d) Number of atoms and the width of the cloud after 10~ms expansion following the release from gray molasses, as a function of b) single photon detuning, c) detuning from the two-photon resonance, and d) cooling time in gray molasses. The initial increase in the number of atoms and the greater uncertainty in the width of the cloud in subplot d) can be attributed to the elevated initial temperature of the atoms: as the cloud expands over 10~ms, a portion of it moves beyond the imaging system's field of view.}
\label{GMC_char}
\end{figure}

Gray optical molasses cooling begins when only $D_1$ light is present (the shutter of the $D_2$ line is completely closed) and the magnetic fields are turned off. At this stage, we detune the $D_1$ line cooling and repumping light by +5$\mathrm{\Gamma}$ from the transition to the excited state $2P_{1/2}~ F'=2$ (Fig. \ref{GMC_char}a) so that a two-photon condition in the $\Lambda$ configuration is satisfied. Figures~\ref{GMC_char}b and \ref{GMC_char}c show the efficiency of gray optical molasses in terms of the remaining atom number and the final temperature as a function of the single- and two-photon detuning from the excited $F'=2$ level.

At the beginning of gray molasess cooling we use the maximum available intensity of the cooling light, $14.6I\mathrm{_s}$, while the intensity of the repumping light ($13.6I\mathrm{_s}$) has been the result of optimization of the transfer of atoms from the D1-D2 CMOT to gray optical molasses. Using AOMs in the amplification module, we linearly reduce the intensity of cooling (repumping) light to 4.6$I\mathrm{_s}$ (4.2$I\mathrm{s}$) in 6~ms (Figure~\ref{GMC_char}d). This procedure decreases the temperature with no apparent loss in sample density, while retaining about 80\% of the initial population. Finally, we obtain a cloud of $\mathrm{4.5\times 10^7}$ atoms, with a temperature of 11~\textmu K (measured using the time-of-flight method) and a phase-space density $\mathrm{\rho = 1.7\times 10^{-5}}$. Such a low temperature is on par with 16~\textmu K reported recently by Cherfan et \textit{al.}~\cite{Cherfan_APL_2021} and nearly 4$\times$ lower than some of the reference results for $^{41}$K~\cite{PhysRevA.84.043432,chen2016production}.

\section{Sub-Doppler cooling and magnetic trapping of the $^{39}$K-$^{41}$K mixture}

The preceding sections have provided a detailed discussion of the laser system and, together with references~\cite{dobosz2021bidirectional} and~\cite{bochenski2024sub}, have established that the reported design can be used for state-of-the-art sub-Doppler cooling of all stable potassium isotopes. We introduce an essential functionality of the laser system: the ability to simultaneously cool a \textit{mixture} of potassium isotopes, which we demonstrate by sub-Doppler cooling and magnetic trapping of a doubly spin-polarized mixture of $^{39}$K-$^{40}$K. 

Among the isotopic mixtures of potassium, Bose-Bose ($^{39}$K-$^{41}$K) and Bose-Fermi ($^{41}$K-$^{40}$K) mixtures have been already brought to quantum degeneracy~\cite{cabrera2018quantumPhD, wu2011strongly}. To the best of our knowledge, simultaneous laser cooling of $^{39}$K and $^{40}$K has not been described in the literature so far. Here, we go beyond showing just that and demonstrate that the mixture can be cooled to sub-Doppler temperatures and confined in a magnetic quadrupole potential with high efficiency. One notable aspect of our finding is the use of a source that contains naturally occurring potassium, where only 0.012\% is $^{40}$K, while almost 93.3\% is $^{39}$K~\cite{tiecke2010properties}. To enhance the number of fermions that can be loaded into the magneto-optical trap, it is necessary to increase the temperature of the dispenser beyond its usual operational parameters. This limits the lifetime of the magnetically trapped mixture to $\sim$1.1~s. 
The number of atoms that can be trapped is constrained by the single-chamber design of the vacuum system and the use of an unenriched atom source. Any increase in atom flux leads to a rise in the background pressure. The output power of the amplifiers utilized in the spectroscopy and amplification modules is another limiting factor. The power of the seeding light must be shared between two isotopes, resulting in a lower power available for each isotope compared to working with a single isotope. These constraints are related to hardware and can be addressed by employing enriched dispensers and more powerful optical amplifiers that are already available commercially.

Because each amplifier is seeded with a minimum of two adjacent frequencies required for two isotopes, it becomes difficult to determine the light intensity for a specific frequency component. In this section of the manuscript, all intensities are estimated based on the power of seeding light and the assumption that for frequencies in close proximity (less than 2~GHz apart), the amplification of each frequency component should be similar. Multifrequency amplification in a tapered amplifier may produce sidebands that could affect the number of atoms or the temperature~\cite{luo2013multiple}. The impact of this phenomenon is minimal, as we have not observed any substantial changes in the cooling efficiency between a single isotope and a mixture. We have also investigated the presence of additional frequency components by monitoring the beatnote between the cooling and the repumping lasers for frequencies differing by up to 500~MHz. The power present in the sidebands observed in the spectrum, which are more than 200~MHz away from the closest transition, has been found to be below 4\%. It should be noted that the existence of sidebands may have a more profound effect on cooling of a $^{39}$K-$^{41}$K mixture, where light frequencies used for cooling (or repumping) two isotopes are much closer to each other.

We start by loading a double isotope magneto-optical trap, which saturates at the $N_{40\mathrm{K}}=3\times 10^5$ and $N_{\mathrm{39K}}=4 \times 10^7$ atoms, with temperatures of $T_{\mathrm{40K}}$=200~\textmu K and ~$T_{\mathrm{39K}}$=1.1~mK, where subscripts indicate the isotope. The optimized parameters for both isotopes can be seen in Table~\ref{detuning}.
\begin{table}[]
\resizebox{\textwidth}{!}{\begin{tabular}{cccccccccc}
\hline
Stage                                                                         & B [G/cm]                                     & Isotope                           & $\mathrm{I_c}$ $\mathrm{[I_s]}$ & $\mathrm{I_r}$ $\mathrm{[I_s]}$ & $\mathrm{\delta_c}$ $\mathrm{[\Gamma]}$ & $\mathrm{\delta_r}$ $\mathrm{[\Gamma]}$ & N                 & T [\textmu K] & $\mathrm{\rho}$       \\ \hline
\multirow{2}{*}{\begin{tabular}[c]{@{}c@{}}MOT\\ (12 s)\end{tabular}}         & \multirow{2}{*}{6.9}                         & $^{39}$K                 & 7.6                             & 3.6                             & -8                                      & -5                                      & $4 \times 10^7$   & 1100          & $7 \times 10^{-10}$   \\
                                                                              &                                              & $^{40}$K                 & 16.5                            & 8                               & -3.8                                    & -1.7                                    & $3 \times 10^5$   & 200           & $2 \times 10^{-10}$   \\ \hline
\multirow{2}{*}{\begin{tabular}[c]{@{}c@{}}CMOT\\  (20 ms)\end{tabular}}      & \multirow{2}{*}{40}                          & $^{39}$K                 & 1.6                             & 1.2                             & -9                                   & -7                                      & $4 \times 10^7$   & 2000          & $8.4 \times 10^{-10}$ \\
                                                                              &                                              & $^{40}$K                 & 19                              & 15                              & -3.5                                    & -3.5                                    & $3 \times 10^5$   & 260           & $2 \times 10^{-9}$    \\ \hline
\multirow{3}{*}{\begin{tabular}[c]{@{}c@{}}D1-D2~CMOT \\ (3 ms)\end{tabular}} & \multirow{3}{*}{$\mathrm{40 \to 0}$} & $^{39}$K                 & $\mathrm{1.2 \to 0}$    & $\mathrm{1.2 \to 0}$    & -8                                      & -4                                      & - & -             & -                     \\
                                                                              &                                              & $^{39}$K$\mathrm{_{D1}}$ & $\mathrm{0 \to 0.4}$    & $\mathrm{0 \to 0.5}$    & -1                                      & +8                                      & -                 & -             & -                     \\
                                                                              &                                              & $^{40}$K                 & $\mathrm{19 \to 0}$     & $\mathrm{5 \to 0 }$     & -3.5                                    & -3.5                                    & -   & -             & -                     \\ \hline
\multirow{2}{*}{\begin{tabular}[c]{@{}c@{}}GMC\\ (12 ms)\end{tabular}}        & \multirow{2}{*}{0}                           & $^{39}$K$\mathrm{_{D1}}$ & 8.8                             & 2.5                            & +4                                    & +13.2                                    & $2 \times 10^7$   & 28            & $1.6 \times 10^{-7}$  \\
                                                                              &                                              & $^{40}$K$\mathrm{_{D1}}$ & 11.4                            & 2.2                             & +28.2                                    & +2.5                                    & $2.5 \times 10^5$ & 13            & $1.2 \times 10^{-7}$  \\ \hline
\end{tabular}}
\caption{\label{detuning} Main cooling stages used in simultaneous sub-Doppler cooling of the $^{39}$K-$^{40}$K mixture and their respective duration. For each stage, we quote the magnetic field gradient, intensities, and detunings of the cooling and repumping light. The intensities are expressed relative to the saturation intensity of the $D_2$ line ($I_{\mathrm{s}}$ = 1.75~mW/cm$^2$), while the detunings from the cooling and repumping transitions are specified relative to the natural width of the $D_2$ line ($\Gamma$= 6.035~MHz). The cloud parameters are not included in the D1-D2~CMOT stage because it is impossible to switch the shutters back to the $D_2$ line for absorption imaging. During the D1-D1 CMOT stage, shutters are switching and $D_1$ and $D_2$ lines are present simultaneously for about 0.6~ms (for $^{39}$K), as elaborated in the main text.}
\end{table}

We have adjusted the intensities and alignment of the cooling beams to enhance the efficiency of sub-Doppler cooling and magnetic trapping. This does not guarantee that the clouds in a magneto-optical trap would overlap well, but ultimately $^{40}$ K is centered within a larger cloud of $^{39}$K. The ratio of the number of fermions ($N_{40}$) to bosons ($N_{39}$) in the MOT is on the order of 1\% and as a result only the fermionic MOT is influenced by the presence of the other isotope. This appears as a decrease in the initial loading rate by 20\% (from $\mathrm{3 \times 10^5 ~\frac{1}{s}}$ to $\mathrm{2.4 \times 10^5 ~s^{-1}}$) and doubling the losses from 0.16~$s^{-1}$ (only ${}^{40}$K present) to 0.30~s$^{-1}$ (${}^{40}$K with ${}^{39}$K present). This could be explained if we assume that the hotter bosonic atoms surrounding the fermionic cloud effectively act like the background gas. Due to the limitations of our imaging system and the small number of fermions involved, we have not been able to perform an analysis that would allow us to quantify the interspecies light induced collisions. Even with increased losses, we observe impressive performance in the following cooling phases, which is comparable to the cooling efficiency of $^{40}$ K alone~\cite{bochenski2024sub}.

\begin{figure}[htbp]
\centering\includegraphics[width=\textwidth]{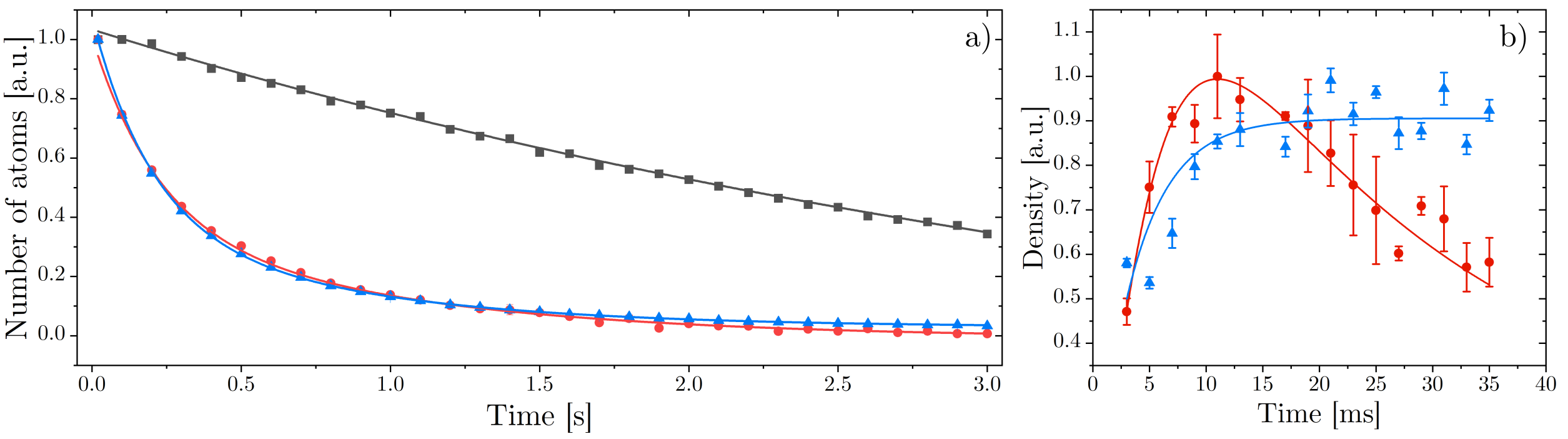}
\caption{a) The evolution of the atom number in a compressed magneto-optical (CMOT) trap over time.
b) Changes in density within the CMOT during a time interval relevant to sub-Doppler cooling. The data points represented by red circles and black squares are for $^{40}$K with and without $^{39}$K present. Blue triangles represent $^{39}$K atoms with $^{40}$K atoms present. The evolution of $^{39}$K atom number with and without $^{40}$K present is indistinguishable, hence the latter scenario is omitted for clarity. The lines are guides to the eye.}
\label{CMOT_char}
\end{figure}

Following the magneto-optical trapping stage, we compress the mixture by increasing the magnetic field gradient and red-detuning the $D_2$ line cooling light (see Table~\ref{detuning} for detailed parameters). We need to find a balance between reaching the highest density of both isotopes and minimizing the loss of atoms due to interspecies and intraspecies collisions. As can be seen in Figure~\ref{CMOT_char}a, losses of both species are rapid. There is no qualitative difference in the case of $^{39}$K compared to a single species operation. However, for $^{40}$K, due to its small density, the difference is notable only in the presence of interspecies loss.

We choose 20~ms as an optimal compression time to ensure consistent daily operation, minimizing the slight dependence of the efficiency of this cooling stage on the alignment of the cooling beams, their powers, and the initial atom number. However, after the chosen compression time, the density of $^{40}$K is typically several percent below its maximum, as can be seen in Figure~\ref{CMOT_char}b, where the maximum density of $n_{\mathrm{40K}} = 9 \times 10^{9}$~$\mathrm{{atoms}/{cm^3}}$ ($n_{\mathrm{39K}} = 1.4 \times 10^{11}$~$\mathrm{{atoms}/{cm^3}}$) for $^{40}$K ($^{39}$K) can be reached after 13~ms (20~ms) of compression. This step increases the density and the phase-space density of both isotopes, with a greater increase of the latter for $^{40}$K (see Table~\ref{detuning}).

A standard compressed MOT (CMOT) is followed by a modified D1-D2 CMOT only for $^{39}$K, similar to the approach chosen with $^{41}$K. For $^{40}$K, whenever cooling is optimized for one of the $D$ lines, the presence of light from the other $D$ line immediately heats the atoms, inducing a large loss. This is purely a consequence of the design of the laser system. Hence, when we transition from the $D_2$ and to the $D_1$ line for $^{40}$K, we ensure that the amplification module does not output light for cooling of the fermionic isotope, as discussed in~\cite{bochenski2024sub}. As a result, there is a 1.2~ms window when the cloud of $^{40}$K expands freely. At the end of this stage the quadrupole field is turned off and we engage the compensation of the stray fields. 

In the final step, we begin simultaneous sub-Doppler cooling in gray optical molasses on the $D_1$ line of each isotope (see Table~\ref{detuning} for experimental details). After 12~ms of cooling we reach the temperature of 28~\textmu K (13~\textmu K) with $2 \times 10^7$ ($2.5 \times 10^5$) atoms at the phase-space density of $1.6 \times 10^{-7}$ ($ 1.2 \times 10^{-7}$) for $^{39}$K ($^{40}$K). At this point, we have lost 50\% bosons and only 15\% fermions. These results may not be as good as those reported in the literature for single-isotope cooling~\cite{salomon2014gray, dobosz2021bidirectional, bochenski2024sub}, but a similar reduction (with respect to single-isotope operation) has also been observed in a recent study of simultaneous sub-Doppler cooling of $^6$Li and $^7$Li~\cite{Dash_6Li_7Li}. Relatively large losses, especially for $^{39}$K, occur during transfer to gray optical molasses. This seems to stem from the limited cooling power in a dual isotope operation, because with the same experimental sequence, but with the laser system set to emit only the $^{39}$K cooling light, the overall performance is nearly identical to what we observe in a single isotope case~\cite{dobosz2021bidirectional}. Choosing not to include a molasses ramping step, commonly used in such experiments, could explain the higher temperatures observed when compared to the temperature of the experiments with a single isotope. The choice to exclude this step has been driven by constraints in laser power, as its inclusion would lead to a significant reduction in the number of atoms. Nevertheless, we have successfully transferred the mixture to a magnetic trap.

We spin polarize both isotopes in two steps. Immediately after gray molasses, we ensure that almost the entire population of each isotope is in a well-defined hyperfine state by performing 100~\textmu s-long hyperfine pumping of $^{39}$K ($^{40}$K) to $^2S_{1/2}$, $F=1$ ($F=7/2$) using $D_1$-line light, where the repumping light for both isotopes has been turned off and only cooling light is present. We then turn on 2.5~G magnetic field in 1~ms to set the quantization axis and turn off the remaining light propagating along the MOT paths. The direction of the magnetic field is experimentally optimized on the basis of the efficiency of the spin polarization. Optical pumping is performed on the $D_1$ line using circularly polarized light, with all beams delivered to the setup via a single polarization-maintaining fiber. Four transitions (two per isotope), each with an intensity of 0.6$I\mathrm{_s}$, are addressed: $2S_{1/2}\,F=1,\,F=2 \to {}^2P_{1/2}\, F=2$ and $^2S_{1/2}\,F=7/2,\,F=9/2 \to {}^2P_{1/2}\, F=9/2$. After 500~\textmu s, beams addressing states with higher $F$ are turned off while the remaining beams are extinguished 200~\textmu s later. This procedure transfers atoms to weak-field seeking states; nearly 90\% of fermions to $F=9/2$, $m_\mathrm{F}$=9/2 without introducing measurable heating, and nearly 80\% of bosons to $F=2$, $m_\mathrm{F}$=2 with a temperature increase by a factor of two. A 57.6~G/cm magnetic field gradient (along the weaker vertical axis) is turned on in 1 ms trapping $1.4\times10^7$  of $^{39}$K and $2.2\times10^5$ of $^{40}$K.

\begin{figure}[htbp]
\centering\includegraphics[width=\textwidth]{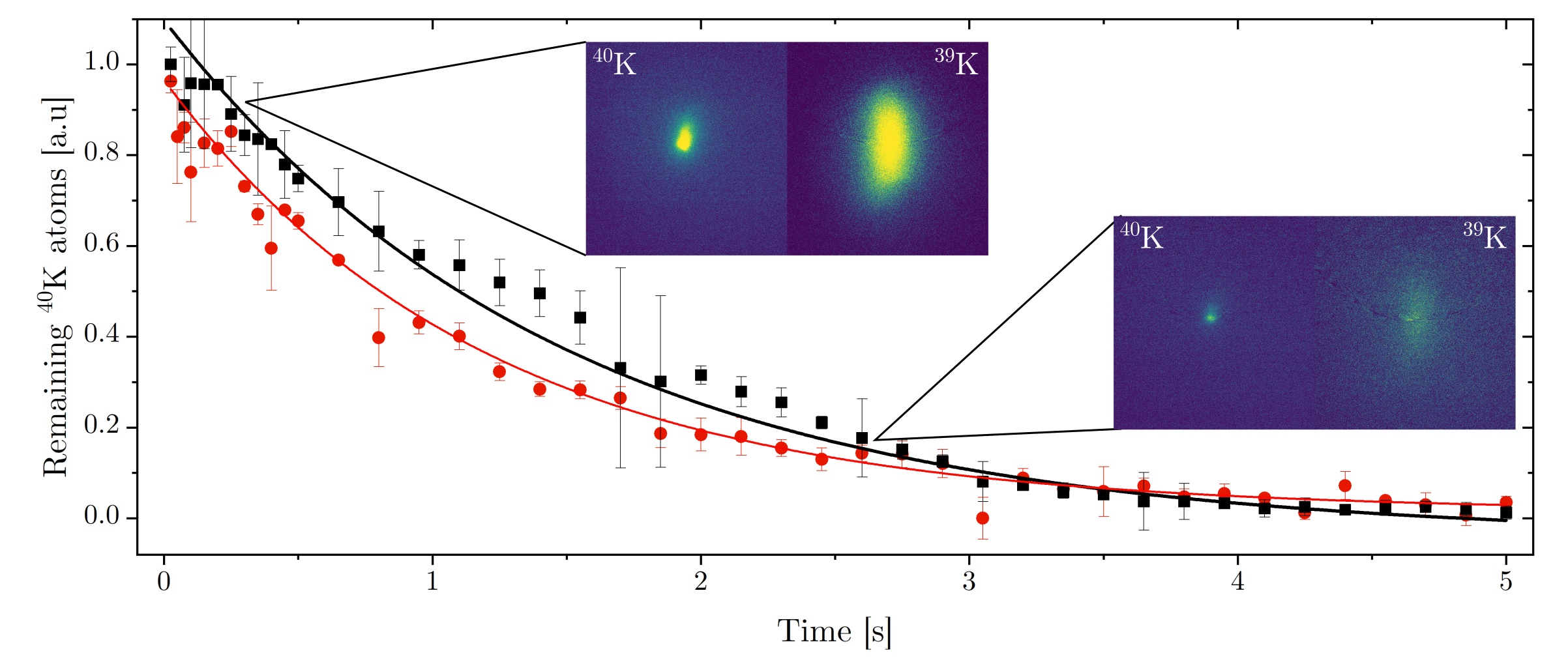}
\caption{The lifetime of $\mathrm{^{40}K}$ atoms in a magnetic trap with a gradient of 57.6 G/cm (aligned with gravity) is examined in two scenarios: a trap containing only a single species (indicated by the red curve, with a lifetime of $\tau = 1500\pm60$~ms) and a trap with the presence of the $^{39}$K isotope (illustrated by the black curve, with a lifetime of $\tau = 1200 \pm 40$~ms). The insets show fluorescence images of both isotopes taken after two different trapping times.}
\label{Decay}
\end{figure}

The measured lifetime of the spin-polarized magnetically trapped atoms (Figure~\ref{Decay}) is limited by the increased background gas pressure, primarily as a result of the operation of the dispensers at an elevated temperature. The $1/e$ lifetimes of fermionic potassium in the single isotope and in the mixture case are $ \mathrm{ \tau = 1500 \pm 60}$ ms and $ \mathrm{ \tau = 1200 \pm 40}$ ms, respectively, with no noticeable difference for $^{39}$ K between these two cases. The results clearly show the appearance of heteronuclear collisions, but increasing the number of fermions or reducing their temperature through rf evaporation would be advantageous for future quantitative studies of the mixture.

\section{Conclusion}

We have successfully created the first, to the best of our knowledge, heteronuclear mixture of $^{39}$K and $^{40}$K trapped in a conservative potential. We have characterized the operation of a dual-isotope magneto-optical and the following cooling stages, including spin polarization of both isotopes. The results are particularly notable because of the simplicity of the experimental setup that uses a single-chamber vacuum system and an unenriched potassium source. The limitations imposed by the lifetime of trapped samples could be overcome if the atoms were magnetically transported to a separate chamber with a much better vacuum. In this particular mixture, evaporative cooling in the magnetic trap is not feasible due to the negative scattering length of $^{39}$K. However, the implementation of degenerate Raman sideband cooling following the gray molasses phase for $^{40}$K, and possibly also for $^{39}$K,~\cite{drsc_39K,drsc_40K} may increase the transfer efficiency into an optical dipole trap such that evaporative cooling optimized for the removal of $^{39}$K might be feasible. However, the demonstrated mixture could be immediately useful for optical tweezer experiments and photoassociation studies of heteroisotopic potassium molecules. The latter could be used to investigate the limits of the Born-Oppenhaimer approximation~\cite{PhysRevA.78.012503,LUTZ201643} by comparison with the spectroscopic data acquired for other homo- and heteronuclear potassium mixtures.

We have demonstrated an innovative laser system design that allows for sub-Doppler cooling of all stable potassium isotopes and their combinations, which has played a crucial role in achieving the results presented in this study. A modular design of the frequency tuning module of the laser system makes it easy to adapt it to cooling radioactive potassium isotopes, for example to study $\beta$ -decay with laser-cooled $^{37}$K and $^{38m}$K isotopes~\cite{behr1997magneto}. A microsecond-scale switching time and an effectively maintenance-free approach to choosing which isotope is cooled make the laser system very useful for experiments in which bosonic and fermionic mixtures~\cite{PhysRevLett.117.163201} or heteronuclear ground-state polar molecules are studied~\cite{PhysRevLett.114.205302,PhysRevLett.125.083401}. In our group, the laser system is currently utilized to investigate properties of ultracold $^{39}$K-Cs and $^{41}$K-Cs mixtures.

The current design of the laser system is relatively expensive, because it utilizes four commercial MOPA systems. One could consider building the spectroscopy module around external cavity diode lasers instead of tapered amplifiers. It would not provide sufficient power ($>$10~mW) for seeding amplifiers in the amplification module, but this could be overcome by using a double-pass of seeding light through the gain medium of the amplifier. This solution has been shown to work for seeding powers as low as 200~\textmu W~\cite{bolpasi2010double} and has been demonstrated for simultaneous trapping of $^{85}$Rb and $^{87}$Rb~\cite{valenzuela2013multiple}.

\begin{backmatter}
\bmsection{Fundings} This research was funded by the Foundation for Polish Science within the Homing program and the National Science Centre of Poland (grant No. 2016/21/D/ST2/02003 and a postdoctoral fellowship for M.S., grant No. DEC-2015/16/S/ST2/00425).
\bmsection{Acknowledgments} We would like to acknowledge P. Arciszewski and K. Din{\c c}er for their contribution to the development of the experimental setup.
\bmsection{Disclosures} The authors declare no conflicts of interest.
\bmsection{Data availability} Data underlying the results presented in this paper are not publicly available at this time but may be obtained from the authors upon reasonable request.
\end{backmatter}

\bibliography{main}

\end{document}